\documentclass{svproc}

\usepackage{url}
\usepackage{graphicx}
\usepackage{amsmath,amssymb}
\usepackage[version=4]{mhchem}
\usepackage{siunitx}
\usepackage{longtable,tabularx}
\usepackage{listings}
\usepackage{caption}

\lstdefinestyle{customc}{
  belowcaptionskip=1\baselineskip,
  breaklines=true,
  frame=L,
  xleftmargin=\parindent,
  language=C,
  showstringspaces=false,
  basicstyle=\footnotesize\ttfamily,
  keywordstyle=\bfseries\color{green!40!black},
  commentstyle=\itshape\color{purple!40!black},
  identifierstyle=\color{blue},
  stringstyle=\color{orange},
}
\newcommand{\RN}{\mathbb{R}}
\newcommand{\dd}{\,\mathrm{d}}
\newcommand{\OF}{J}
\newcommand{\OFC}{\tilde J}
\newcommand{\FACE}{\mathcal{S}}
\newcommand{\GALL}{\mathcal{G}}
\newcommand{\RALL}{\mathcal{R}}
\newcommand{\UALL}{U}
\newcommand{\LALL}{\lambda}
\newcommand{\CC}{C\nolinebreak\hspace{-.05em}\raisebox{.4ex}{\tiny\bf +}\nolinebreak\hspace{-.10em}\raisebox{.4ex}{\tiny\bf +}}
\def\CC{{C\nolinebreak[4]\hspace{-.05em}\raisebox{.4ex}{\tiny\bf ++}}}

\begin{document}
\mainmatter
\title{Accurate gradient computations for\\ shape optimization via discrete
adjoints in CFD-related multiphysics problems}
\titlerunning{Discrete Adjoints for Multiphysics}
\author{Ole Burghardt \and Nicolas R. Gauger}
\tocauthor{Ole Burghardt, Nicolas R. Gauger}
\institute{TU Kaiserslautern, Chair for Scientific Computing,\\ 67663 Kaiserslautern, Germany,\\
\email{ole.burghardt@scicomp.uni-kl.de},\\ WWW home page:
\texttt{https://www.scicomp.uni-kl.de/}}
\maketitle
\begin{abstract}
As more and more multiphysics effects are entering the field of CFD simulations, this raises the question how they can be accurately captured in gradient computations for shape optimization.\\
The latter has been successfully enriched over the last years by the use of (discrete) adjoints. One can think of them as Lagrange multipliers to the flow field problem linked to an objective function that depends on quantities like pressure or momentums, and they will set also the framework for this paper.\\
It is split into two main parts: First, we show how one can compute coupled discrete adjoints using automatic differentiation in an effective way that is still easily extendable for all kinds of other couplings.\\ 
Second, we suppose that a valuable first application are so-called conjugate heat transfer problems which are gaining more and more interest from the automobile and aeronautics industry. Therefore we present an implementation for this capability within the open-source solver SU2 \cite{economon2015su2} as well as for the generic adjoint computation algorithm.
\end{abstract}
\section{Introduction}
Two quite different research fields have emerged to compute the adjoint vector field - commonly referred to as $\lambda$.\\ 
Whereas in the first one a partial differential equation in terms of $\lambda$ is set up and subsequently discretized and solved (continuous approach), the other approach directly solves for $\LALL$ within the discretized formulation (discrete approach).\\

Studies which method is preferable over the other are carried out along questions of computational effort, accuracy and stability issues.\\ 
While the discrete adjoint method has shown significant advantages in the latter two cases, it should be emphasized that the replacement itself of the adjoint PDE model (which has to be determined in the first place) by techniques like automatic differentiation (AD) creates the opportunity for a very generic implementation which can be particularly exploited when it comes to multiphysics problems and gradient computations therein.
\subsection{Mathematical formulation of CFD solvers}
As we will be working in the discretized framework, we can directly introduce CFD solvers rather than a PDE model (which could be the Euler equations just as well as the Reynolds-averaged Navier-Stokes equations) as the main object we are looking at. We denote them by a function
\begin{equation}\label{CFDSolver}
\GALL_{(X)}:\RN^{mn} \rightarrow \RN^{mn} \\
\end{equation}
mapping the intermediate solution $\UALL_i$ to $\UALL_{i+1}$, where $X \in \RN^{m}$ is denoting the (computational mesh) coordinates and $n$ is the number of components (formerly the number equations of the original PDE problem).\\
We assume that the solution $\UALL^*(X)$ to the PDE problem is given by the fixed point of $\GALL$ (SU2 and almost any other CFD solver is working in this manner).\\ On the computational side, it is the first intermediate fulfilling a convergence criteria like
\begin{equation*}
\| \GALL_{(X)}(\UALL_i) - \UALL_i \| < \varepsilon,
\end{equation*}
still, for all mathematical considerations, we assume that $\GALL_{(X)}(\UALL^*)=\UALL^*$.\\

When evaluating an objective function $J$ based on $\UALL^*$, its value can as well depend explicitly on the geometry $X$ but is, moreover, only depended on $X$ as the CFD solver does not change during a shape optimization run and a flow solution is given implicitly by the fixed point of $\GALL_{(X)}$.\\
We denote this by defining an objective function $\OFC:\RN^{m} \times \RN^{mn}\rightarrow\RN$ and setting 
\begin{equation*}
\OF(x) := \OFC(x,u(x))
\end{equation*}  
accordingly.
This implicit dependence turns out to be a severe problem if we now want to compute the gradient of $\OF$ with respect to $X$,
\begin{equation*}
D\OF (X) = \frac{\partial}{\partial x}\OFC(X,\UALL(X)) + \frac{\partial}{\partial u}\OFC(X,\UALL(X))\cdot \frac{\partial}{\partial x}\UALL(X),
\end{equation*}
as the last factor would involve the complex program $\GALL_{(X)}$ and would be too expensive to compute for large $m$. Here is where adjoint solutions will step in, and we will address their theory in the next chapter.
\subsection{Rewriting $J(X)$ as a Lagrangian}
The effect of changing flow field variables under a (small) variation of $X$ can be efficiently captured by a Lagrange multiplier.\\ 
Let us introduce the function $\tilde\GALL: \RN^{m}\times \RN^{mn}\rightarrow \RN^{mn}$ which is, in constrast to $\GALL$, explicitly dependent on the computational mesh $X\in\RN^{m}$.\\
 
Now set up the Lagrangian $L:\RN^{m}\times \RN^{mn}\rightarrow \RN$ by defining
\begin{equation*}
L(x,u) := \OFC(x,u) + (\tilde\GALL(x,u) - u)^T\cdot \LALL.
\end{equation*}
Restricted to the set $S\subset\RN^{mn}$ of actual flow solutions, $L$ equals $\OFC$, independent of the choice of $\LALL$. In particular, $D\OF (X)=\frac{\dd}{\dd x}L(X,\UALL(X))$.\\
As usual, let $\LALL$ be the multiplier so that
\begin{equation}\label{AdjointEquation}
\nabla_u \OFC (X,U) \stackrel{!}{=} -D_u\RALL^T(X,\UALL)\cdot\LALL,
\end{equation}
where $\RALL(x,u):=\GALL(x,u) - u$. (Which is nothing but the claim that $\frac{\partial}{\partial u} L = 0$.)\\
If we now knew $\LALL$, $DJ(X)$ would be easy to compute as no terms $\frac{\partial}{\partial x} U(X)$ appear anymore, that is
\begin{equation}\label{GradientJ}
D\OF (X)=\frac{\partial}{\partial x}L(X,\UALL(X))=\frac{\partial}{\partial x}\OFC(X,U)+\frac{\partial}{\partial x}\RALL^T(X,U)\cdot\lambda.
\end{equation}
\subsection{Automatic differentiation in reverse mode}
In \cite{Albring_etal2016b}, the authors successfully developed a discrete adjoint solver in SU2 for aerodynamic shape optimizations where they implemented the fixed-point iteration equivalent for \eqref{AdjointEquation},
\begin{equation}\label{iterL}
\LALL \stackrel{!}{=} \nabla_u \OFC(X,U) + D_u\GALL^T (X,U)\cdot\LALL.
\end{equation}
Here, for each new iterate $\LALL_{i+1}$, one needs to evaluate $D_u\GALL^T (X,U)\cdot\LALL_i$, which happens to fit the formulation of the reverse mode of AD.\\ 
There are many tools available, but in \cite{Albring_etal2016b}, CoDiPack \cite{sagebaum2017high} was chosen for its performance and convenience to use within $\CC$ codes.\\ We will go more into detail on how it works (to some extent) and how it is applied once we introduced the specific structure of $\GALL$ for multiphysics.
\section{Coupled simulations}\label{chap:CoupledSimulations}
Multiphysics solvers actually consist of different iterators $\GALL^{(1)},\GALL^{(2)},..., \GALL^{(r)}$ that are only valid in specific parts of the geometry, often referred to as zones (or their computational meshes $X^{(1)},X^{(2)},...,X^{(r)}$).\\ At the interfaces of these zones, solution values or derived ones are exchanged in every iteration step and a solution is given by the combination of the fixed-point solutions $\UALL^{(1)},\UALL^{(2)},...,\UALL^{(r)}$ from each and every zone, meaning that also all interface data agrees.\\ 
Still, by regarding all iterators $\GALL^{(1)},\GALL^{(2)},..., \GALL^{(r)}$ as components of one $\GALL$, we abstractly keep the structure that we introduced in the context of one CFD solver.\\
The mapping in \eqref{CFDSolver} adapts to\\
\begin{equation}\label{coupledPrimal}
\UALL^{(k)}_i \mapsto \GALL^{(k)}_{(X^k,\UALL^{(1)}_i,...,\UALL^{(k-1)}_i,\UALL^{(k+1)}_i,\UALL^{(r)}_i)}(\UALL^{(k)}_i) = \UALL^{(k)}_{i+1},
\end{equation}
where $k$ refers to the zones.\\
For sequential couplings, an outline of \eqref{coupledPrimal} is given by
\begin{lstlisting}[caption={Outline for a multiphysics driver.},captionpos=b,label=code:MultiphysicsDriver1]
while(!convergence) {
  for(k = 0; k < nZones; k++) {
    for(j = 0; j < InnerIter[k]; j++) {
      U[k]->SetSolution(G(k,U));
    }    
  }
  OuterIteration++;
}
\end{lstlisting}
where we added the possibility to drive zone-dependent inner iterations. We leave them out for all theory explanations (as we already did in \eqref{coupledPrimal}) as they are purely a matter of performance and stability and are not needed for gradient computations like $D_u \GALL$ later on.
\subsection{Basic terminology of automatic differentiation}
In order to demonstrate the adjoint version of \ref{code:MultiphysicsDriver1}, this sub-chapter introduces some notation and vocabulary for AD that is taken from \cite{sagebaum2017high}.\\
From their perspective, a computer program is a map $f:\RN^a\rightarrow\RN^b$ that can be represented as a sequence of $l$ \emph{statements} $\varphi_i:\RN^{n_i}\rightarrow\RN$.\\
For each $i$, we refer to $v_i$ as the output values in the image of $\varphi_i$. Its input values are denoted by $w_i$ which would be just a vector of $n_i$ preceding output values, denoted by
\begin{equation*}
w_i := (v_j)_{j\prec i} \in \RN^{n_i}.
\end{equation*}
Evaluating $f$ with respect to some input values $u_i\in\RN^a$ and output values $y_i\in\RN^b$ can now be described as
\begin{center}
\begin{tabular}{ p{1.5cm} p{5cm} p{2cm} }
  $v_i$ & $=u_i$ & $i=1\dots a$ \\
  $v_{i+a}$ & $=\varphi_i(w_i)$ & $i=1\dots l$ \\
  $y_i$ & $=v_{a+l-i+1}$ & $i=1\dots b$. \\
\end{tabular}
\end{center}
We now regard our iterator $\GALL$ as such a function $f$. If during its evaluation the derivative information of all its statements could be stored, that is the values
\begin{equation}\label{ADDerivatives}
\frac{\partial}{\partial v_j}\varphi_i(w_i),
\end{equation}
we could then obtain all intermediate derivatives with respect to the $v_i$ (indicated by bars and commonly simply called ``the adjoint values'') and a arbitrary but fixed $\bar\lambda\in \RN^b$ by executing\\
\begin{center}
\begin{tabular}{ p{1.5cm} p{5cm} p{2cm} }
  $\bar v_{a+l-i+1}$ & $=\bar y_i$ & $i=b\dots 1$ \\
  $\bar v_j$ & $=\bar v_j + \bar v_{i+n}\cdot\frac{\partial}{\partial v_j}\varphi_i(w_i), \bar v_i=0$ & $i=l\dots 1$ \\
  $\bar u_i$ & $=\bar v_i$ & $i=a\dots 1$,\\
\end{tabular}
\captionof{table}{Reverse mode run.}
\label{table:ReverseMode} 
\end{center}
giving
\begin{equation}\label{ReverseDerivative}
\bar u = D_u\GALL^T(X,U)\cdot\bar\lambda,
\end{equation}
which is exactly what we need for \eqref{iterL}.\\
The information \eqref{ADDerivatives} is stored in a \emph{tape} during a dedicated run of $\GALL$ (at given solution $U$). For referencing it and to link it to the adjoint values, it is internally assigned \emph{indices}, incremented for each and every statement being executed.\\
To have access to the desired derivatives, all input values have to be \emph{registered} which will assign them indices before running the program. The same accounts for the output values to indicate where the adjoint values $\bar\lambda$ have to be applied.\\

In multiphysics, where we will have multiple sets of input and output values (above all, the solution variables in the different zones), this internal referencing of variables will turn out to be the part where one has to be careful, as explained next.
\subsection{Reverse mode applied to multiphysics drivers}
The simplest solution would be to register all input sets from the different zones, record the tape of a fully coupled run and update all adjoint solutions in all zones afterwards. In regard of \eqref{iterL}, this translates into performing
\begin{equation*}\label{updateL}
\begin{pmatrix}
\LALL^{(1)}_{i+1} \\
\LALL^{(2)}_{i+1} \\
\vdots \\
\LALL^{(r)}_{i+1}
\end{pmatrix}
=
\begin{pmatrix}
\nabla_{u^{(1)}}\tilde J(U) \\
\nabla_{u^{(2)}}\tilde J(U) \\
\vdots \\
\nabla_{u^{(r)}}\tilde J(U)
\end{pmatrix}
+
\begin{pmatrix}
\frac{\partial}{\partial u^{(1)}}\GALL^{(1)}(U) & \dots & \dots & \frac{\partial}{\partial u^{(1)}}\GALL^{(r)}(U) \\
\frac{\partial}{\partial u^{(2)}}\GALL^{(1)}(U) & \ddots & & \frac{\partial}{\partial u^{(r)}}\GALL^{(1)}(U) \\
\vdots & & & \vdots\\
\frac{\partial}{\partial u^{(r)}}\GALL^{(1)}(U) & \dots & \dots & \frac{\partial}{\partial u^{(r)}}\GALL^{(r)}(U)
\end{pmatrix}\cdot
\begin{pmatrix}
\LALL^{(1)}_{i} \\
\LALL^{(2)}_{i} \\
\vdots \\
\LALL^{(r)}_{i}
\end{pmatrix}
\end{equation*}
where the cross terms arise from the dependencies of solutions from other zones, as indicated in \eqref{coupledPrimal}.\\
Unfortunately, this (direct) approach comes with some drawbacks by its unmodular design, the most important ones being
\begin{itemize}
\item The impossibility for driving zone-wise inner iterations,
\item limited possibility to intentionally neglect (unstable) dependencies,
\item the need for special treatments of certain parts in the primal code which are error-prone and should be independent from the adjoint code development.
\end{itemize}
The last point will become particularly important for coupled iterators. An ad-hoc implementation of their taping for a 2-zone problem would look like
\begin{lstlisting}[caption={Ad-hoc taping routine for $\GALL$.},captionpos=b,label=code:AdhocTape]
StartRecording();
RegisterInput(U[0]); RegisterInput(U[1]);
U[0]->SetSolution(G(0,U)); RegisterOutput(U[0]);
U[1]->SetSolution(G(1,U)); RegisterOutput(U[1]);
StopRecording();
\end{lstlisting}
giving -- for example -- wrong results for $\frac{\partial}{\partial u^{(1)}} \GALL^{(2)}$ as the solution vector $\UALL^{(1)}$ has been updated though we would have wanted to extract the derivative with respect to the original solution set to capture the coupling.\\ 
One can think of various workarounds within the primal code to prohibit such behaviour or even to re-tape before each and every iteration. Though especially the latter is not favorable for performance reasons.\\

A more promising approach is to directly work with the internal indices of the AD tool which will be just a further \texttt{int} data structure within the different solvers. A sketch is shown in \ref{code:IndexTape}.
\begin{lstlisting}[caption={Index-based taping routine for $\GALL$.},captionpos=b,label=code:IndexTape]
StartRecording(); PushBackTapePosition();
for(k = 0; k < nZones; k++) {
  RegisterInput(U[k]); 
  SetIndices(InputIndices[k],U[k]);
}
PushBackTapePosition();
for(k = 0; k < nZones; k++) {
  U[k]->SetSolution(G(k,U));
  SetIndices(OutputIndices[k],U[k]);
  PushBackTapePosition();
}
StopRecording();
\end{lstlisting}
The evaluation counterpart is given by \ref{code:EvaluateL} below.
\begin{lstlisting}[caption={Part of the adjoint solution \texttt{L} update (the contribution from $J$ being left out which would be just an initialization with $1$).},captionpos=b,label=code:EvaluateL]
L_Old.SetSolution(L);
for(k = 0; k < nZones; k++) {
  SetAdjoints(OutputIndices[k],L_Old[k]);
  Tape.Evaluate(k+2,k+1); Tape.Evaluate(1,0);
  for(j = 0; j < nZones; j++) {
    L[j].SetSolution(ExtractAdjoints(InputIndices[j]);
  }
  L_Iter.AddSolution(L);
}
L.SetSolution(L_Iter);
\end{lstlisting}
Tape positions are saved just in order to keep the possibility to only evaluate parts of it later what would be the case if one would like to do multiple updates with respect to specific adjoint value sets.\\
Its exhaustive version has been implemented in SU2 to allow for multiphysics discrete adjoints without the need to add or change problem-specific code once a new functionality has been added on the primal side.
\section{A multiphysics key example: conjugate heat transfer}
\label{chap:CHT}
Conjugate heat transfer (in short CHT) problems are of special interest when it comes to optimizations of devices where the conservation of energy across all physical zones cannot be neglected. Think, for example, of cooling devices or turbine blades in a high-temperature airflow.
\subsection{Heat equation discretization}
SU2 started as an aerodynamics flow solver but by its modular way of implementation it is easily extendable so that it evolved to cover incompressible regimes as well as multiphysics capacities for FSI and turbomachinery, too.\\ An CHT implementation requires two components: a solid heat solver and an appropriate coupling routine. In solid domains $\Omega$, the static energy equation simplifies to covering heat conduction only, that is
\begin{eqnarray}\label{Laplace}
\frac{k}{\rho c_p}\Delta T(x) = 0 & \text{for all} & x\in\Omega,
\end{eqnarray}
subject to prescribed boundary conditions, which will be solved by a finite-volume approach. To fit it into the pseudo-time formulation in SU2 (following an idea of Chorin \cite{Chorin}) to obtain -- in regard to \eqref{coupledPrimal} -- a similar iterative scheme as for the flow solver, we add  an artificial time dependence to it, to then obtain the steady-state limit of
\begin{equation}\label{HeatEquation}
\frac{\partial}{\partial \tau} T(t,x) - \frac{k}{\rho c_p}\Delta T(t,x) = 0,
\end{equation}
where $k$ is the thermal conductivity which we assume to be a constant, $\rho$ its density, $c_p$ its specific heat capacity and $\tau$ the artificial time (meaning that $\frac{\partial}{\partial\tau} T = 0$ implies $\frac{k}{\rho c_p}\Delta T = 0$).\\
To discretize, we require \eqref{HeatEquation} to hold true for every cell $V\subset\Omega$ within a computational mesh (with $n$ being its outer normal), that is
\begin{eqnarray}\label{HeatEquationIntegrated}
\int\limits_V \frac{\partial}{\partial \tau} T(t,x) \dd x = \int\limits_V \frac{k}{\rho c_p} \Delta T(t,x) \dd x = \int\limits_{\partial V} \frac{k}{\rho c_p}\nabla T(t,x)\cdot n \dd S.  
\end{eqnarray}
To approximate the projected heat flux $\frac{k}{\rho c_p}\nabla T\cdot n$ on every face $\FACE_i\subset \partial V$, we approximate $(\nabla T\cdot n)|_{\FACE_i}$ by taking a finite difference of the temperature values at its adjacent nodes $T_i$ and $T_j$ (connected by some vector $\nu$) and correct it in the same manner as proposed in \cite{economon2015su2} using results from \cite{Weiss}, giving\\
\begin{equation}
(\nabla T\cdot n)|_{\FACE_i} = \frac{T_j-T_i}{\| \nu \|}(\nu\cdot n) + \frac{1}{2}(\nabla T_i + \nabla T_j)(n-(\nu\cdot n)n),
\end{equation}
where $\nabla T_i$ and $\nabla T_j$ are the approximated gradient values at the nodes, e.g. obtained by a Green-Gauss method.\\
The time derivative in \eqref{HeatEquationIntegrated} is discretized by a finite difference in time (the integral being approximated by a multiplication with $|V|$) and we could obtain a temperature solution by running an explicit time stepping.\\ 
For an implicit Newton-like solver, we need the Jacobian for the right hand side of \eqref{HeatEquationIntegrated} with respect to the temperatures at the respective nodes, which we approximate to equal $-\frac{k}{\rho c_p}(\nu\cdot n)$ at $T_i$ and $\frac{k}{\rho c_p}(\nu\cdot n)$ at $T_j$, respectively.
\subsection{Heat transfer between fluids and solids}
Our assumption for running a simulation that possibly consist of fluid and solid zones which share common interfaces is that there exists a temperature distribution at their interface that gives the same heat fluxes on both sides, also with respect to differing heat conductivities.\\
In regard to \eqref{coupledPrimal}, one could explicitly derive it for each and every node in the interface from both solver's current temperature solutions (the coupled one and the one performing the update) and set it as a (strong) Dirichlet boundary condition on both sides. This turns out to be unstable in agreement to the theoretical stability analysis on that topic, e.g. in \cite{Giles}.\\
An alternative is to interpret the solid's temperature as the fluid zone's interface temperature and to evoke the heat fluxes to agree by setting the ones obtained in the fluid zone as a (weak) Neumann or Robin boundary condition on the solid zone's interface.\\
In short, for every node at position $x$ on the interface (and $c(x):\frac{k}{d(x)}$, where $d(x)$ denotes the distance to the nearest node in normal direction), we have
\begin{eqnarray}
\begin{cases}
T_i^F(x) = T_i^S(x) & \text{at all fluid-sided interface nodes}\\
& \text{and} \\
h_i^S(x) = c^F(x) \cdot (T_i^S(x)-T_n^F(x) ) & \text{at all solid-sided interface nodes}\\
& \text{or} \\
h_i^S(x) = c^F(x) \cdot (T_i^F(x)-T_n^F(x) ) & \text{at all solid-sided interface nodes}.
\end{cases}
\end{eqnarray}
All options are implemented as different boundary condition routines being executed after all data appearing on the right hand side has been communicated through generic transfer routines.\\

\subsection{Gradient validation}
\begin{figure}
\centering
\begin{minipage}[t]{.45\textwidth}
\centering
\includegraphics[height=4.5cm]{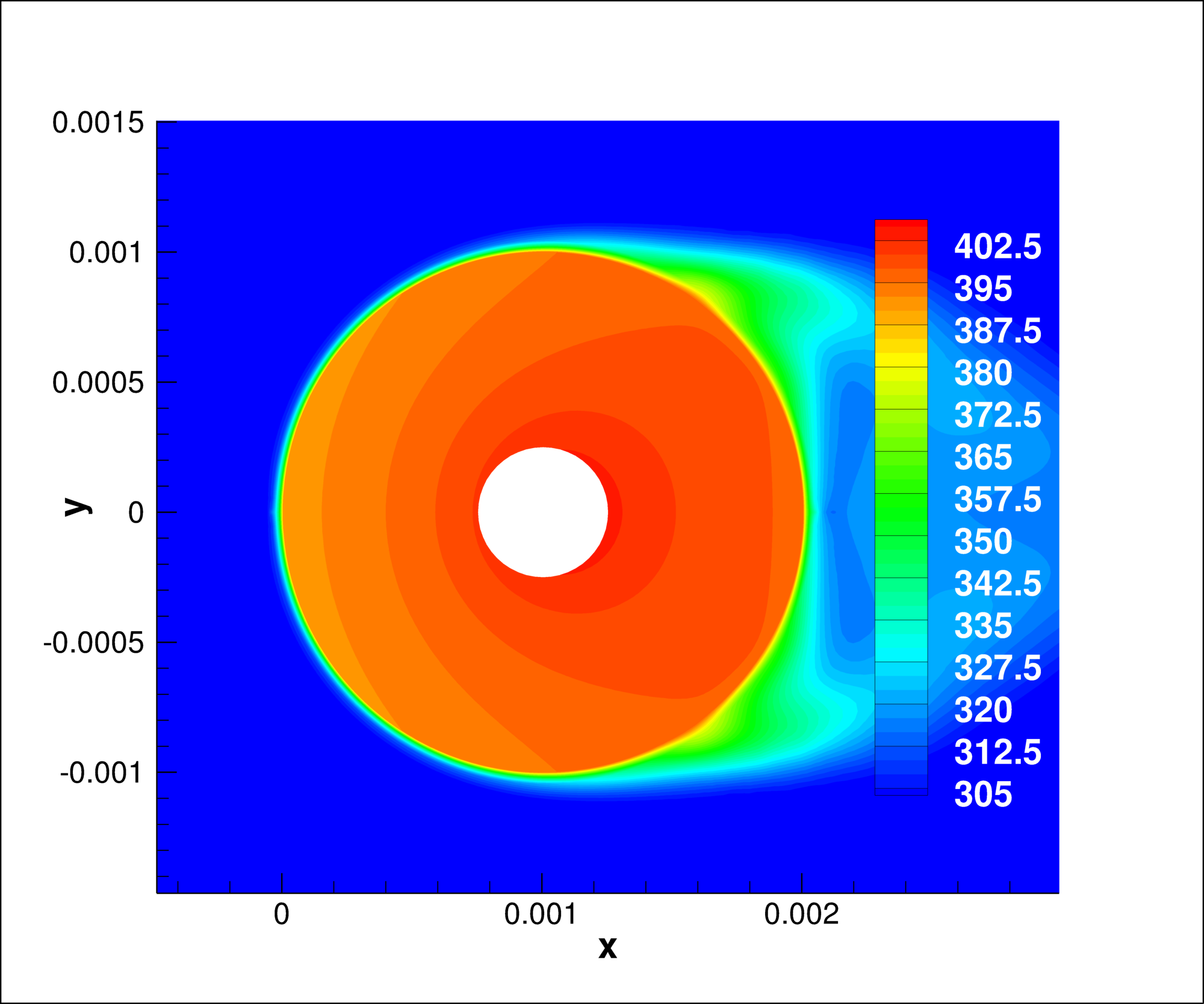}
\caption{Heated cylinder in fluid flow. Note the better cooling effect in upstream direction. The averaged temperature at the inner perimeter is set as $J$. Initially we obtain $J=402.04K$.}
\label{fig:pin}
\end{minipage}%
\hspace{10pt}
\begin{minipage}[t]{.45\textwidth}
\centering
\includegraphics[height=4.5cm]{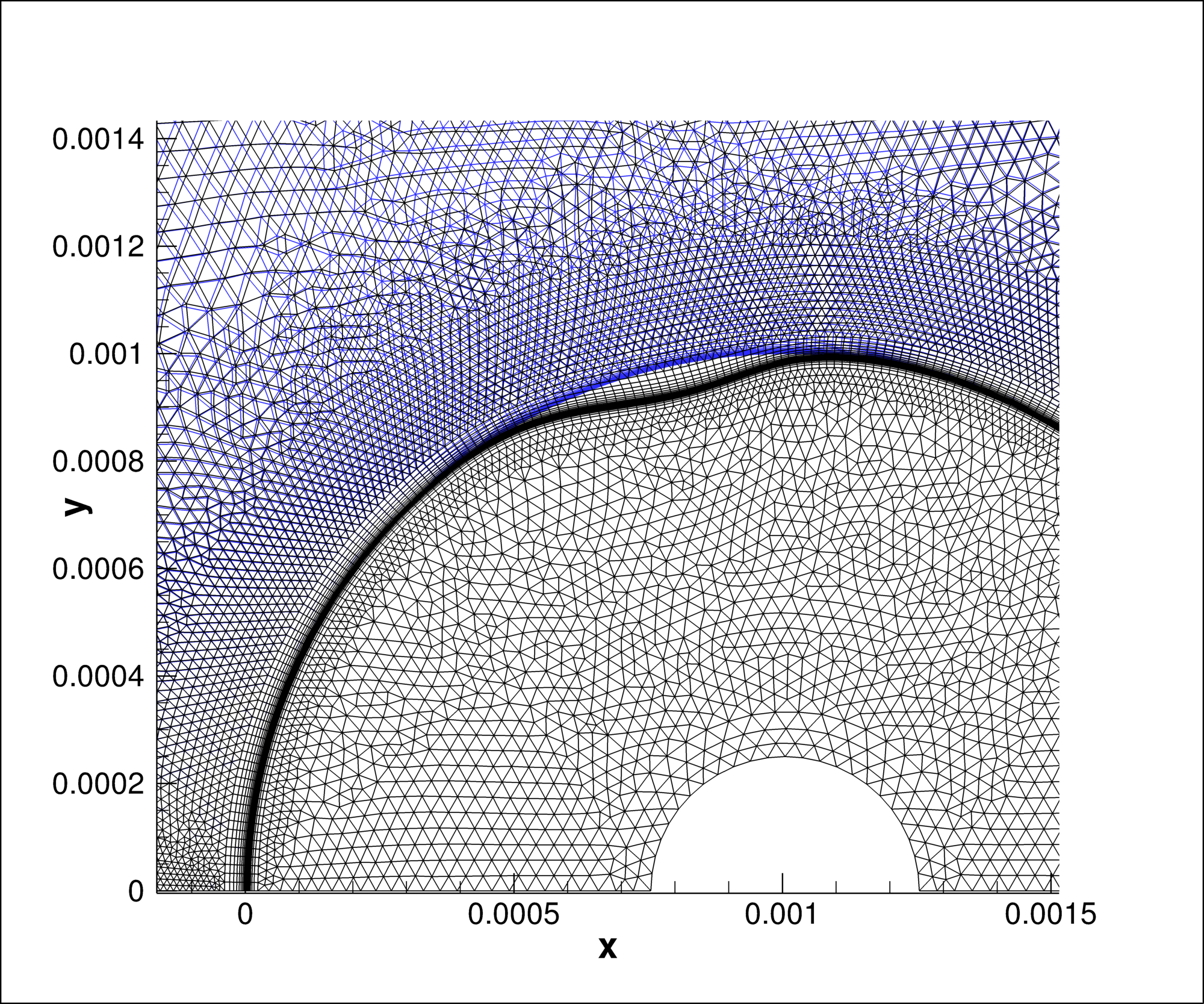}
\caption{Visualization of $\delta$. It moves all grid node positions so that we can compare the change of $J$ with our obtained gradient $\nabla J(X)$ by evaluating it in direction $\delta$.}
\label{fig:meshdef}
\end{minipage}
\end{figure}

To validate shape gradients obtained by the gradient computation method presented above for the CHT implementation, we set up a test case that originates from designing high-performance pin-fin coolers (chosen because of their simplicity in terms of their problem setting). Here, solid zones subject to a fixed heat load transfer all heat into a coolant fluid through some parts of their surfaces. The temperature they reach thereby is to be found by simulations.\\
In particular, we created a 2-d aluminium hollow cylinder which is surrounded by some water flow entering from the left at $0.25\frac{m}{s}$ and $300K$ and which is heated by $4\frac{kW}{m}$ at the inner perimeter. The simulation result being shown in \ref{fig:pin}.\\
%\vspace{-10pm}

\begin{figure}
\centering
\includegraphics[height=4.5cm]{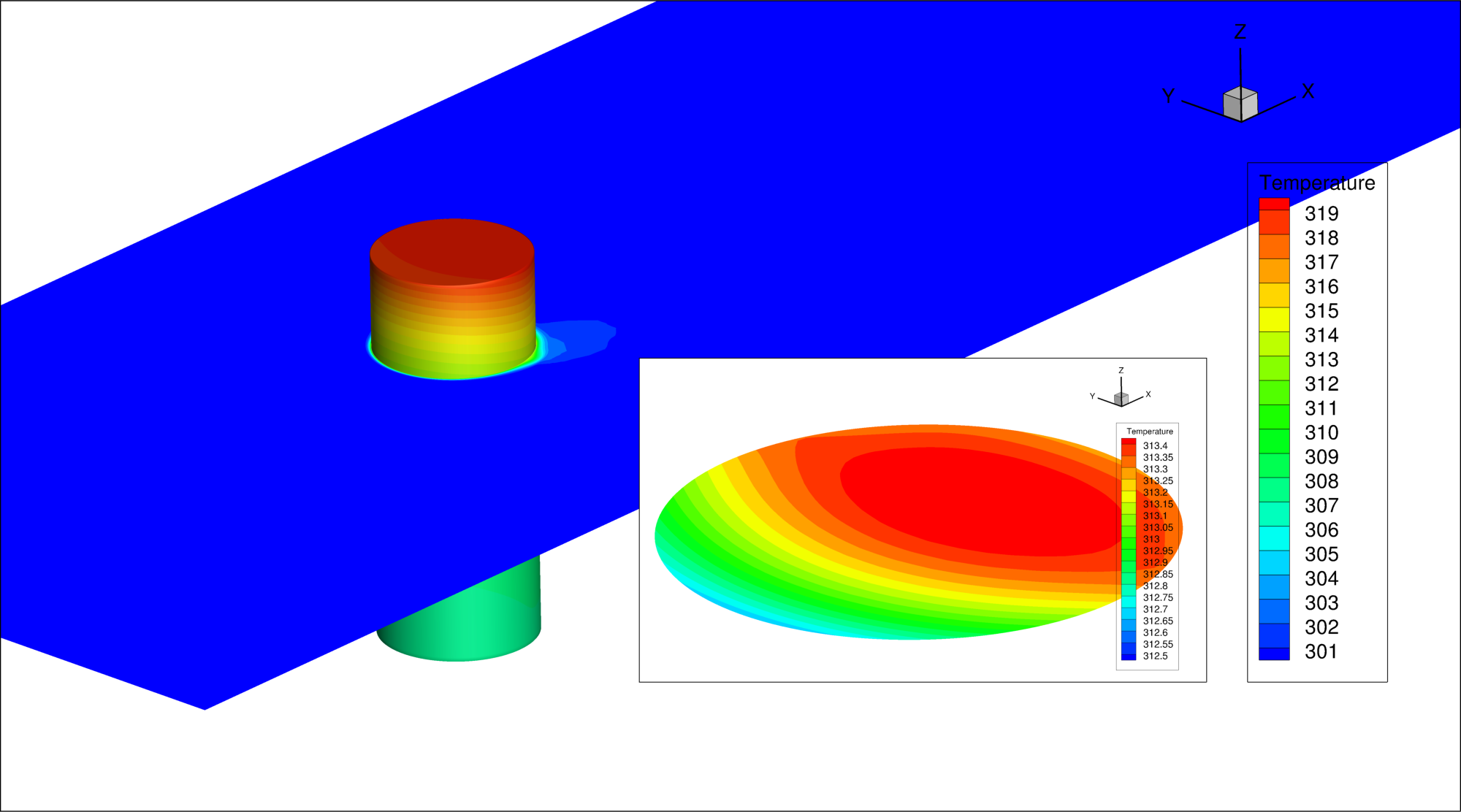}
\caption{Heated cylinder in 3d. Here, the pin is heated (by $4W$) only at its tip and we notice a temperature increase also in $z$-direction, revealing an eligible space for gradient-driven shape optimization.}
\label{fig:3d}
\end{figure}

Applying the coupled discrete adjoint implementation now gives the adjoint vector fields $\lambda^{(1)}$ and $\lambda^{(2)}$ for the flow and the solid zone, respectively. Applying formula \eqref{GradientJ} gives $\nabla J(X)$.\\ 
Let us now introduce a small variation $\delta$ of the underlying computational mesh coordinates $X$ (which we denote by $X_h:=X+h\delta$, graphically shown in \ref{fig:meshdef}) to check in terms of finite differences whether $\nabla J(X)$ is correct, that is whether 
$\lim\limits_{h\rightarrow 0} \frac{J(X_h)-J(X)}{h} = \nabla J(X)\cdot\delta$
%\begin{equation}\label{FiniteDifference}
%\lim\limits_{h\rightarrow 0} \frac{J(X_h)-J(X)}{h} = \nabla J(X)\cdot\delta
%\end{equation}
holds true.\\
%For a magnitude of $h=10^{-6}m$ the quotient gives $-3062.01\frac{K}{m}$, representing an relative error of $46,68\%$ to the expected value, but already for $h=10^{-7}m$ the error improves to $0.01\%$ (with the quotient giving $-5743.35\frac{K}{m}$).\\ 
Already for $h=10^{-7}m$ the relative error between the expected value and the quotient (giving $-5743.35\frac{K}{m}$) reduces to $0.01\%$.\\  
This agreement proves our gradients to be correct and we could use them without any difficulties for shape optimizations, also in 3 dimensions (the counterpart for the validation case being shown in \ref{fig:3d}) and for arbitrary complex geometries as we made no assumptions on the flow solver.\\ 
We intend to publish results in this regard soon. Especially for optimizations of turbine blades where temperatures appear as constraints during the design process, exact gradients could be of high interest.\\
\section{Conclusion}
In this paper, we introduced an algorithm for computing coupled discrete adjoints for multiphysics optimization problems by using AD in reverse mode that is efficient und does not rely on problem-specific code. Using the structures for AD that were already available, we then implemented it in SU2.\\ 
Further, we added a CHT functionality being a multiphysics application where exact gradients gained high interest and
 proved our adjoint solutions to be exact by validating the derived gradients against finite differences.
%\paragraph{Notes and Comments.}
%The actual porgram code following the outlines presented in chapter \ref{chap:CoupledSimulations} and %\ref{chap:CHT} is free to use and to change.\\
%It can be accessed on \texttt{https://github.com/su2code/SU2}. For further information on SU2 see %\texttt{https://su2code.github.io}. Please be aware that some (recent) developments may take their time to %enter the stable master version. In case you are missing what you are looking for, feel free to contact me via %the email address mentioned above.
%


\begin{thebibliography}{6}
%

\bibitem {economon2015su2}
Economon, T. D., Palacios, F., Copeland, S. R., Lukaczyk, T. W., Alonso, J. J.: SU2: an open-source suite for multiphysics simulation and design.
AIAA Journal, vol. 54, pp. 828-846 (2015).

\bibitem {Albring_etal2016b}
Albring, T., Sagebaum, M., Gauger, N. R.: Efficient Aerodynamic Design using the Discrete Adjoint Method in SU2.
AIAA 2016-3518 (2016).

\bibitem {sagebaum2017high}
Sagebaum, M., Albring, T., Gauger, N. R.: High-Performance Derivative Computations using CoDiPack.
arXiv preprint arXiv:1709.07229 (2017).

\bibitem {Chorin}
Chorin, A.: A numerical method for solving incompressible viscous flow problems.
Journal of Computational Physics, vol. 2, pp. 12-26 (1967).

\bibitem {Weiss}
Weiss, J. M., Maruszewski, J. P., Wayne, A. S.: Implicit solution of the Navier-Stokes equation on unstructured meshes.
AIAA Journal, 97-2103 (1997).

\bibitem {Giles}
Giles, M.-B.: Implicit solution of the Navier-Stokes equation on unstructured meshes.
AIAA Journal, 97-2103 (1997).
\end{thebibliography}
\end{document}